\documentclass[fleqn,10pt]{wlscirep}
\title{Selective mass enhancement close to the quantum critical point in 
B\MakeLowercase{a}F\MakeLowercase{e$_2$}(A\MakeLowercase{s$_{1-x}$}P\MakeLowercase{$_x$)$_2$}}

\author[1,2,3]{V. Grinenko}
\author[2,3]{K. Iida}
\author[1,2]{F. Kurth}
\author[2]{D. V. Efremov}
\author[2]{S.-L. Drechsler}
\author[4]{I. Cherniavskii}
\author[2,4]{I. Morozov}
\author[2,5]{J. H\"{a}nisch}
\author[6]{T. F\"{o}rster}
\author[7]{C. Tarantini}
\author[7]{J. Jaroszynski}
\author[8]{B. Maiorov}
\author[8]{M. Jaime}
\author[9]{A. Yamamoto}
\author[3]{I. Nakamura}
\author[3]{R. Fujimoto}
\author[3]{T. Hatano}
\author[3]{H. Ikuta}
\author[2]{R. H\"{u}hne}

\affil[1]{Institute for Solid State Physics, TU Dresden, 01069 Dresden, Germany}
\affil[2]{IFW Dresden, Helmholtzstrasse 20, 01069 Dresden, Germany} 
\affil[3]{Department of Crystalline Materials Science, Graduate School of Engineering, Nagoya University, Furo-cho, Chikusa-ku, Nagoya 464-8603, Japan}
\affil[4]{Lomonosov Moscow State University, GSP-1, Leninskie Gory, Moscow, 119991, Russian Federation}
\affil[5]{Karlsruhe Institute of Technology, Institute for Technical Physics, Hermann-von-Helmholtz-Platz 1, 76344 Eggenstein-Leopoldshafen, Germany}
\affil[6]{Hochfeld-Magnetlabor Dresden (HLD-EMFL), Helmholtz-Zentrum Dresden-Rossendorf, 01314 Dresden, Germany}
\affil[7]{NHMFL, Florida State University, Tallahassee, FL 32310, USA}
\affil[8]{MPA-CMMS, Los Alamos National Laboratory, Los Alamos, NM, 87545, USA}
\affil[9]{Department of Applied Physics, Tokyo University of Agriculture and Technology 2-24-16 Nakacho, Koganei, Tokyo 184-8588, Japan}


\begin{abstract}
A quantum critical point (QCP) is currently  being conjectured for the BaFe$_2$(As$_{1-x}$P$_x$)$_2$ system at the critical value $x_{\rm c} \approx$ 0.3. In the proximity of a QCP, all thermodynamic and transport properties are expected to scale with a single characteristic energy, given by the quantum fluctuations. Such an universal behavior has not, however, been found in the superconducting upper critical field $H_{\rm c2}$. Here we report $H_{\rm c2}$-data for epitaxial thin films extracted from the electrical resistance measured in very high magnetic fields up to 67 Tesla. Using a multi-band analysis we find that $H_{\rm c2}$ is sensitive to the QCP, implying a significant charge carrier effective mass enhancement at the doping-induced QCP that is essentially band-dependent. Our results point to two qualitatively different groups of electrons in BaFe$_2$(As$_{1-x}$P$_x$)$_2$. The first one (possibly associated to hot spots or whole Fermi sheets) has a strong mass enhancement at the QCP, and the second one is insensitive to the QCP. The observed duality could also be present in many other quantum critical systems. 
\end{abstract}
\begin{document}

\flushbottom
\maketitle
%
%
\thispagestyle{empty}

\section*{Introduction}

In most of unconventional superconductors, a quantum critical point (QCP) of 
charge or spin density wave (CDW/SDW) states lies beneath the superconducting dome.\cite{Scalapino2012,Hirschfeld2011,Gabovich2002,Gegenwart2008} 
Low-energy quantum fluctuations in the vicinity of a QCP lead to non-Femi 
liquid (nFL) behavior in the normal state and a strong enhancement of the 
effective electron mass ($m^*$). A good example is given by heavy fermion 
superconductors. In some of these systems the maximum superconducting transition 
temperature ($T_{\rm c}$) coincides with the position of the expected QCP of the 
magnetic phase.\cite{Gegenwart2008} The presence of a QCP beneath the 
superconducting dome is evidenced by a strong enhancement of the superconducting 
specific heat jump $\Delta C/T_{\rm c}$ at $T_{\rm c}$ and the slope of the upper critical field $|H_{\rm c2}'| = |{\rm d}H_{\rm c2}/{\rm d}T|$ normalized by the critical temperature in the vicinity of $T_{\rm c}$.\cite{Bauer2010}

\begin{figure}
	\centering
	\includegraphics[width=9.1cm]{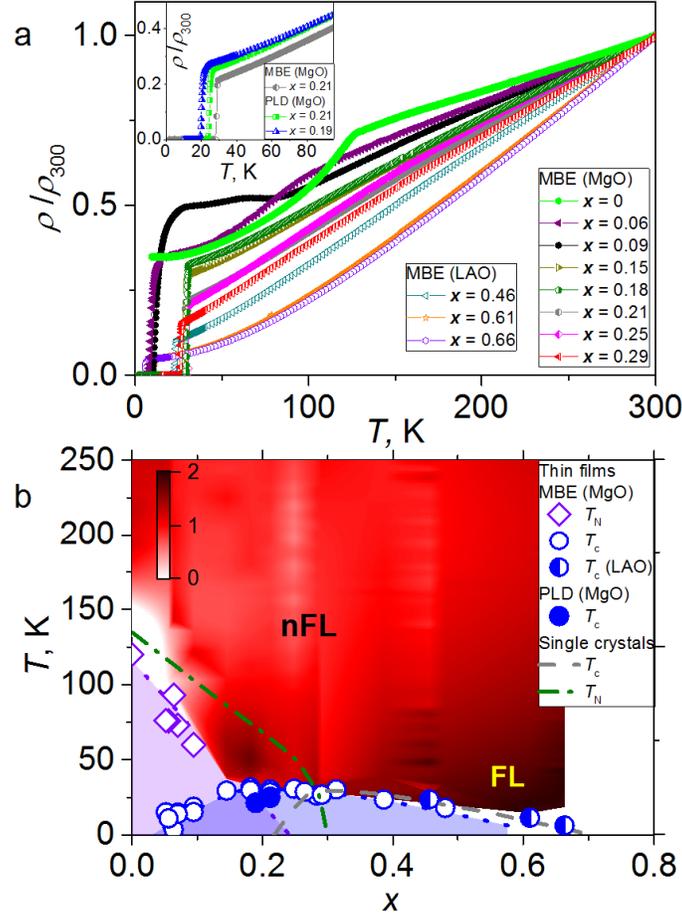}
		\caption{({\bf a}) The temperature dependence of the normalized resistivity $\rho/\rho_{300K}$ of BaFe$_2$(As$_{1-x}$P$_x$)$_2$ films prepared by MBE. Closed symbols - underdoped, half closed symbols - optimally doped, and open symbols - overdoped samples. Inset: The normalized resistivity traces for BaFe$_2$(As$_{1-x}$P$_x$)$_2$ thin films with the similar P-doping prepared by PLD and MBE. ({\bf b}) The phase diagram of BaFe$_2$(As$_{1-x}$P$_x$)$_2$ thin films (symbols). The data of BaFe$_2$(As$_{1-x}$P$_x$)$_2$ single crystals (dashed lines) \cite{Shibauchi2014,Kasahara2010} are also shown for comparison. The whole 
phase diagram for the thin films prepared on MgO substrates is shifted to lower doping levels compared to that of the single crystals and films prepared on LAO substrates. The shift of the phase diagram is substrate-dependent due to different in-plane strain. The contour plot of the doping and temperature dependence of the exponent $n$ is obtained from the data shown in Fig.\ref{Fig:1}a and in Ref. \cite{Kawaguchi2014} assuming  $\rho$ = $\rho_0$ + $AT^n$. The position of the QCP is around $x_{\rm c} \sim 0.25$ for the films prepared on MgO substrates. The positions of the QCP for the single crystals and films prepared on LAO substrates are nearly coincides at $x_{\rm c} \sim 0.30$. For further details see text. The doted lines are guides to the eyes.}
\label{Fig:1}
\end{figure}

In multi-band iron-based superconductors (FBS), the maximum of $T_{\rm c}$ is 
usually linked to the expected position of a QCP of the SDW 
phase.\cite{Johnston2010} Evidence for a zero-temperature second order magnetic 
transition with pronounced quantum fluctuations was found for optimally doped 
BaFe$_2$(As$_{1-x}$P$_x$)$_2$ by various measurements in the normal 
state.\cite{Shishido2010,Analytis2014,Hayes2016,Nakai2010,Iye2012,Nakai2013} 
Therefore, it is considered to be a classical example of unconventional 
superconductivity emerging in the vicinity of a magnetic 
state.\cite{Abrahams2010,Shibauchi2014} However, no doping dependence of the 
scattering rates expected for a QCP scenario was observed in recent 
angle-resolved photoemission spectroscopy (ARPES) studies.\cite{Fink2015} 
In the superconducting state, a divergent quasiparticle effective mass ($m^*$)
above the QCP of the SDW phase was suggested based on specific heat 
\cite{Walmsley2013} and penetration depth measurements 
\cite{Hashimoto2012,Lamhot2015} as well as predicted by theoretical 
studies.\cite{Levchenko2013,Nomoto2013} However, $H_{\rm c2}$ at low $T$ and its slope near $T_{\rm c}$ are insensitive to the QCP.\cite{Putzke2014} This behavior is seemingly in contradiction to many other experimental observations. To resolve this puzzle we investigated in detail the temperature dependence of $H_{\rm c2}$ for BaFe$_2$(As$_{1-x}$P$_x$)$_2$ single-crystalline thin films in a wide range of P-doping. The obtained data can be described in an effective two-band model with qualitatively different doping dependences of the Fermi velocities ($v_{\rm F}$). Namely, $v_{\rm F1}$ is indeed nearly featureless across the QCP implying a doping independent $m_1^*$. On the other hand, $v_{\rm F2}$ is strongly doping-dependent, in accord with the almost divergent logarithmic enhancement of $m_2^*$ observed in many other experiments.

\section*{Results}

\subsection*{Electronic phase diagram of BaFe$_2$(As$_{1-x}$P$_x$)$_2$}
BaFe$_2$(As$_{1-x}$P$_x$)$_2$ epitaxial thin films were grown by molecular beam 
epitaxy (MBE).\cite{Kurth2015,Kawaguchi2014} The investigated MBE thin films 
have high crystalline quality with $T_{\rm c}$ values above 30 K at optimal 
doping level. Some of the films were prepared by pulsed laser deposition 
(PLD). The PLD films have slightly reduced 
$T_{\rm c}$ at similar doping levels compared to the films prepared by MBE as shown in inset of Fig.\ \ref{Fig:1}a. This result is consistent with previous 
studies.\cite{Sato2014} To construct the phase diagram of our thin films, we 
analyzed the temperature dependence of the resistivity for various 
doping levels shown in Fig.\ \ref{Fig:1}a. The 
phase diagrams of the BaFe$_2$(As$_{1-x}$P$_x$)$_2$ thin films and 
single crystals \cite{Shibauchi2014,Kasahara2010} are shown in Fig.\ \ref{Fig:1}b. The whole 
phase diagram for the thin films prepared on MgO substrates is shifted to lower doping levels compared to that of the single crystals. The shift of the phase diagram, as it was shown in previous studies, is substrate-dependent due to different in-plane 
strain.\cite{Kurth2015,Kawaguchi2014,Iida2009,Engelmann2013,Iida2016} In particular, the  in-plane tensile strain for the films grown on MgO modifies slightly the position of the bands resulting in the observed difference between the phase diagrams of thin films and single crystals.\cite{Iida2016} On the other hand, the amount of strain for the films grown on LaAlO$_3$ (LAO) is negligibly small resulting in the same phase diagram as for single crystals.\cite{Kawaguchi2014}

\begin{figure}
	\centering
	\includegraphics[width=9.1cm]{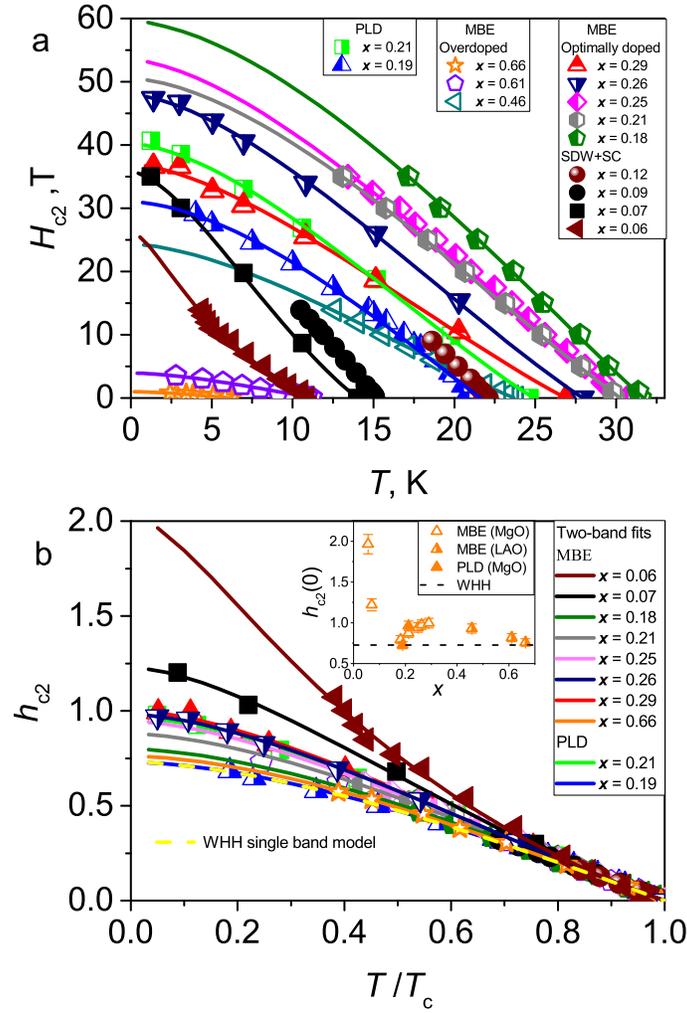}
		\caption{({\bf a}) Temperature dependences of the upper critical field $H_{\rm c2}$ of BaFe$_2$(As$_{1-x}$P$_x$)$_2$ thin films with various doping levels for the magnetic field applied along the $c$-axis. Closed symbols - underdoped, half-open symbols - optimally doped, and open symbols - overdoped samples,  solid lines - two-band fits. ({\bf b})  The reduced field $h_{\rm c2} = \frac{H_{\rm c2}}{-H_{\rm c2}'T_{\rm c}}$ as a function of $T/T_{\rm c}$ for the data shown in Fig. \ref{Fig:2}a, solid lines - two-band fits (the same as in Fig. \ref{Fig:2}a), dashed line - single-band WHH model. The inset shows the doping dependence of the $h_{\rm c2}(0)$ values extrapolated to $T = 0$. The deviation of the experimental data from the single-band curve indicates a relevance of multi-band effects for the temperature dependencies of $H_{\rm c2}$. This deviation is doping dependent as shown in the inset and it can be described by a two-band model for a clean superconductor with a dominate interband coupling. For further details see text.}
\label{Fig:2}
\end{figure}

We assumed that the temperature dependences of the resistivity (Fig.\ \ref{Fig:1}a) can be described by $\rho$ = $\rho_0$ + $AT^n$ in the normal state above the superconducting and magnetic transition temperatures. This general expression has been frequently employed in the quantum critical region, where $n$ = 1 at the QCP and $n$ = 2 in a Fermi liquid (FL) state.\cite{Analytis2014,Shibauchi2014} The contour plot in Fig.\ \ref{Fig:1}b illustrates the temperature and doping dependences of the exponent $n = \frac{Td^2\rho/dT^2}{d\rho/dT} +1$, as 
calculated using experimental temperature dependences of the resistivity. In this analysis we exclude the data close to the SDW transition, where $d\rho/dT \lesssim 0$ (white region in Fig.\ \ref{Fig:1}b).
The region in the phase diagram with nFL behavior is similar to the single 
crystals: the exponent $n$ shows a V-shape; however, it shifts to lower doping 
level. This allows to estimate the critical doping level for thin films on MgO substrates as $x_{\rm c} \approx 0.25\pm 0.03$, which is slightly lower than $x_{\rm c} \approx 0.3$ reported for single crystals.\cite{Shibauchi2014} For the films prepared on LAO substrate we assumed that the position of the QCP coincides with the QCP position for the single crystals due to close similarity between their phase diagrams as discussed above.

\subsection*{Upper critical field} 

The temperature dependences of $H_{\rm c2}$ for BaFe$_2$(As$_{1-x}$P$_x$)$_2$ thin films with various doping levels for fields parallel to the $c$-axis are shown in Fig.\ \ref{Fig:2}. The temperature dependence of $H_{\rm c2}$ is strongly affected by the amount of doping. To compare the data of samples with different doping levels, we plot the reduced field 
$h_{\rm c2} = \frac{H_{\rm c2}}{-H_{\rm c2}'T_{\rm c}}$ versus the reduced 
temperature $t = T/T_{\rm c}$ in Fig.\ \ref{Fig:2}b, where 
$H_{\rm c2}'$ is the extrapolated slope of $H_{\rm c2}$ at $T_{\rm c}$. For the strongly overdoped, and slightly underdoped samples, $0.15 < x < 0.21$, the experimental $h_{\rm c2}$ data are close to the prediction of 
the single-band Werthamer-Helfand-Hohenberg (WHH) model which includes only the 
orbital pair-breaking effect.\cite{Werthamer1966} For other 
doping levels, the experimental $h_{\rm c2}$ data deviate from the single band fit. The doping dependence of $h_{\rm c2}(0)$ extrapolated to zero temperature is shown in the inset of Fig.\ \ref{Fig:2}b. 
The $h_{\rm c2}(0)$ values exhibit a broad maximum around optimal doping $x_{\rm c}$. Additionally, $h_{\rm c2}(0)$ is strongly enhanced in the coexistence state between SC and magnetism, where $T_{\rm N} > T_{\rm c}$.

The doping evolution of the temperature dependences of $H_{\rm c2}$ can be described by the two-band model for a clean superconductor as proposed by Gurevich \cite{Gurevich2010,Gurevich2011} assuming a dominant interband coupling $\lambda_{12}\lambda_{21}\gg\lambda_{11}\lambda_{22}$ as expected for $s_{\pm}$ superconductors. The expression for $B\parallel c$ is given in the Supplementary material Eq.\ S1. A small value of the intraband couplings $\lambda_{11} = \lambda_{22} \sim 0.1$ affects the resulting Fermi velocities within 10 \%, only around optimal doping (see Fig.\ S7) and it has a negligible effect for overdoped samples. Therefore, to reduce the number of fitting parameters, we adopted zero intraband $\lambda_{11} = \lambda_{22} = 0$ pairing constants. In this case, the superconducting transition temperature is related to the coupling constants by  $T_{\rm c} = 1.14\Omega_{\rm sf} {\rm e}^{(-1/\lambda_{12}\lambda_{21})}$. We considered two different values of the characteristic spin fluctuation energy $\Omega_{\rm sf}$, 100 K and 62 K, in order to take into account a possible softening of the spin fluctuations spectrum at the QCP. We assumed also that the paramagnetic pair breaking is negligibly weak, $\alpha_{\rm M} << 1$, as suggested by the small electronic susceptibility of BaFe$_2$(As$_{1-x}$P$_x$)$_2$, where the Maki parameter $\alpha_{\rm M} = 2^{1/2} H_{\rm c2}^{\rm orb}/H_{\rm p}$, defined by a ratio between orbital critical field $H_{\rm c2}^{\rm orb}$ and Pauli limiting field $H_{\rm p}$,  quantifies the strength of the paramagnetic pair breaking (see also the Supplementary material). This assumption is consistent with a relatively small Knight shift of BaFe$_2$(As$_{1-x}$P$_x$)$_2$.\cite{Nakai2013} The result of the fit is shown in Fig. \ref{Fig:2}, and the obtained fitting parameters are given in the Supplementary tables (Tabs. S1 and S2).

\section*{Discussion}

\begin{figure}
	\centering
	\includegraphics[width=9.1cm]{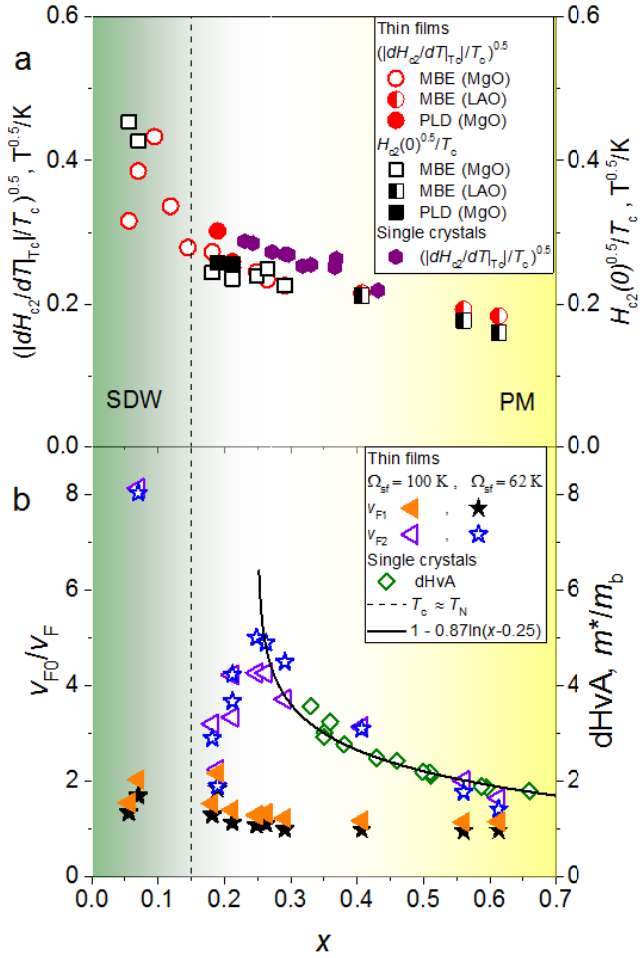}
		\caption{({\bf a}, left axis) The normalized slope of the upper critical field $(|H_{\rm c2}'|/T_{\rm c})^{0.5}$ at $T_{\rm c}$ and ({\bf a}, right axis) the normalized upper critical field $(H_{\rm c2}(0))^{0.5}/T_{\rm c}$ extrapolated to $T = 0$ using the fits shown in Fig.\ \ref{Fig:2} versus the P-doping level $x$. Both these quantities are related to the charge carrier effective mass $m^*$ as discussed in the text. The single crystalline data are taken from Ref. \cite{Putzke2014}. ({\bf b}, left axis) The inversed normalized Fermi velocities $v_{\rm F0}/v_{{\rm F}1}$ and $v_{\rm F0}/v_{{\rm F}2}$ in a two-band model are obtained from the fits shown in Fig.\ \ref{Fig:2}. ({\bf b}, right axis) The normalized effective quasiparticle mass $m^*/m_{\rm b}$ obtained from the dHvA data.\cite{Shishido2010,Walmsley2013} The $v_{\rm F0}$ values are chosen to fit  dHvA data, $v_{\rm F0} = 1.3 \cdot 10^{7}$ and $1.1 \cdot 10^{7}$ cm s$^{-1}$ for $\Omega_{\rm sf} = 100$ K and  62 K, respectively. All the data of the thin films grown on LAO substrate, and single crystalline data are shifted by $\Delta x$ = - 0.05 to meet the QCP of the thin films grown on MgO substrate according to Fig.\ \ref{Fig:1}b. The solid line is a fit to a phenomenological divergence of the effective mass near a QCP $1/v_{\rm F1} \propto m^* \propto 1+0.87{\rm ln}(x-x_{\rm c})$ Refs. \cite{Analytis2014,Abrahams2010} with $x_{\rm c}$ = 0.25.  The dashed line denotes approximately the P doping value where $T_{\rm c} = T_{\rm N}$.}
\label{Fig:3}
\end{figure}

The doping dependencies of $|H_{\rm c2}'/T_{\rm c}|^{0.5}$ extrapolated to 
$T_{\rm c}$, and the $H_{\rm c2}(0)^{0.5}/T_{\rm c}$ extrapolated to $T = 0$ are shown in Fig. \ref{Fig:3}a. According to the BCS theory for clean superconductors, these values are proportional to the quasiparticle effective mass ($m^*$). As it was pointed out in Ref. \cite{Putzke2014}, $|H_{\rm c2}'/T_{\rm c}|^{0.5}$ should have a peak-like maximum at the QCP of the SDW phase since $m^*$ is strongly enhanced near optimal doping on the whole Fermi surface according to various experimental data.\cite{Shishido2010,Walmsley2013,Hashimoto2012} However, this is not the case: $|H_{\rm c2}'/T_{\rm c}|^{0.5}$ and $H_{\rm c2}(0)^{0.5}/T_{\rm c}$ are nearly featureless at optimal doping ($x_{\rm c} \sim$ 0.25) in accord with Ref. \cite{Putzke2014}. Both the single crystals and our MBE films have high $T_{\rm c}$ values of about 30 K at optimal doping indicating similar low impurity scattering rates. The slightly higher $|H_{\rm c2}'/T_{\rm c}|^{0.5}$ values of 
the single crystals compared to those of the MBE films are probably related to the different experimental methods used for the evaluation of $H_{\rm c2}$. 
Also, $H_{\rm c2}$ of the PLD films follows the same trend in spite of a lower $T_{\rm c}$ and residual resistivity ratio (inset 
of Fig. \ref{Fig:1}a) as compared to those of the MBE films. Therefore, we believe  that the observed doping dependence of $H_{\rm c2}$ is not affected essentially by impurity scattering rates and related instead mainly to the changes of $v_{\rm F}$ and the coupling constants.       

$H_{\rm c2}$ of multi-band unconventional $s$-wave superconductors with dominant interband coupling is limited by the largest $v_{\rm F}$ in the usually considered pronounced $s_{\pm}$-regime.\cite{Gurevich2010,Gurevich2011} Therefore, in the case of a strong 
$v_{\rm F}$ asymmetry between different bands, the larger Fermi velocity ($v_{\rm F1}$ in our notation) dominates $H_{\rm c2}$ around optimal doping. In this case one can write $(H_{\rm c2}'/T_{\rm c})^{0.5}\propto H_{\rm c2}(0)^{0.5}/T_{\rm c} \propto v_{\rm F1}^{-1}\propto m_1^*$. This explains the observed weak doping dependence of these quantities (Fig.\ \ref{Fig:3}). The obtained doping dependences of the (normalized reciprocal) $v_{\rm F1}$ and $v_{\rm F2}$ are shown in Fig. \ref{Fig:3}b. The $1/v_{\rm F1}$ values are indeed smaller than $1/v_{\rm F2}$ and show a weaker doping dependence. In contrast, $1/v_{\rm F2}$ is strongly enhanced around optimal doping. The $\Omega_{\rm sf}$ value affects the Fermi velocities quantitatively but their qualitative doping dependence is conserved. The corresponding normalized effective mass $ m^*/m_{\rm b}$ obtained from the de Haas-van Alphen (dHvA) experiments \cite{Shishido2010,Walmsley2013} follows the same trend if the small shift of the QCP along the doping axis ($\Delta x$ = -0.05) due to the strain is taken into account (see Fig. \ref{Fig:1}b), where $m_{\rm b}$ is the quasiparticle mass taken from the band structure calculations. The logarithmic divergence at $x_{\rm c}$ = 0.25 is an indication for the reduction of $v_{\rm F2}$ caused by the quantum fluctuations associated with a QCP of the SDW phase.\cite{Analytis2014,Walmsley2013} A strong reduction of $v_{\rm F2}$ is observed also at $x < 0.15$ which roughly corresponds to the doping level where $T_{\rm N} > T_{\rm c}$ (Fig.\ref{Fig:1}b, see also Tabs.\ S1 and S2 in the Supplementary material). This behavior may be associated with the reconstruction of the Fermi surface due to the presence of the coexisting SDW phase.\cite{Fink2015,Yi2012,Nakashima2013} 

Some of the multi-band heavy fermion superconductors show a similar behavior around the magnetic QCP as compared to the BaFe$_2$(As$_{1-x}$P$_x$)$_2$ system. 
The measured enhancement of the effective mass depends also essentially on the 
experimental method.\cite{Knebel2008} Also, a seemingly conflicting behavior between the dHvA, ARPES and transport data was discussed for cuprate superconductors around optimal hole doping.\cite{Senthil2014} It was proposed that for the suggested nodal electron pocket induced by bidirectional charge order in high fields, the mass enhancement is very anisotropic around the small Fermi surface. It was argued that the corners of that pocket exhibit a large enhancement without any enhancement along the diagonal nodal direction. Such an angle-dependent mass enhancement is interpreted as a destruction of the Landau quasiparticles at 'hot spots' on the large Fermi surface at a proximate QCP. Moreover, another recent theoretical work questioned the paradigm of the universal nFL behavior at a QCP.\cite{Paul2016} It was shown that at the nematic QCP the thermodynamics may remain of FL type, while, depending on the Fermi surface geometry, either the entire Fermi surface stays cold, or at most there are 'hot spots'. Therefore, one may speculate that the complex behavior observed in FBS and in particular for BaFe$_2$(As$_{1-x}$P$_x$)$_2$ can be related to the superposition of {\it two distinct} QCPs associated with the SDW phase and the nematic order.\cite{Chowdhury2015} The evidence for two distinct QCPs was indeed reported for the Ba(Fe$_{1-x}$Ni$_x$)$_2$As$_2$ system.\cite{Zhou2013} Recently, a band-dependent mass enhancement toward the QCP was suggested from the high-field specific heat measurements of overdoped BaFe$_2$(As$_{1-x}$P$_x$)$_2$ single crystals.\cite{Moir2016} Thus far, the available experimental data emphasize the relevance of multi-band effects for a proper and complete understanding of the quantum criticality of BaFe$_2$(As$_{1-x}$P$_x$)$_2$ and related systems. Further theoretical and experimental investigations would be helpful to develop a microscopic scenario of the QCP for the title compound and other multi-band systems. 

\section*{Methods}

\subsection*{Samples}
BaFe$_2$(As$_{1-x}$P$_x$)$_2$ single crystalline thin films with various  P 
doping levels $x$ were grown by MBE with a background pressure of the order of 
10$^{-7}$ Pa. All elements were supplied from solid sources charged in Knudsen 
cells. Pure elements were used as sources for Ba, Fe, and As. The P$_2$ flux was 
supplied from a GaP decomposition source where Ga was removed by two trapping 
caps placed on the crucible. The details of the sample preparation are given in 
Refs. \cite{Kurth2015,Kawaguchi2014}. 
Some of the films on MgO (100) substrate were prepared by PLD with a KrF excimer laser (248 nm). In this case, we used polycrystalline BaFe$_2$(As$_{1-x}$P$_x$)$_2$ as 
the PLD target material. The preparation process took place in an ultra-high vacuum 
chamber with a similar base pressure of 10$^{-7}$ Pa. Before the deposition, the 
substrate was heated to 850 $^{\rm o}$C. Then the BaFe$_2$(As$_{1-x}$P$_x$)$_2$ 
layer was grown with a laser repetition rate of 3 Hz. The layer thickness was 
adjusted via the pulse number at constant laser energy. To improve the sample's 
homogeneity and thickness gradient, the substrate was rotated during the whole 
deposition process. Phase purity and crystalline quality of the films were examined by X-ray diffraction (XRD). The $c$-axis lattice parameters were calculated from the XRD data using the Nelson Riley function. It depends linearly on the P-doping (determined by electron probe micro-analysis (EPMA)) for the films grown on the same 
substrate.\cite{Kawaguchi2014} In this work, we mainly investigated films prepared on MgO (100) substrate. At high doping levels, also several films on LaAlO$_3$ (100) substrate have been used. The P-doping levels given in the paper have been determined using the $c$-axis lattice parameter values according to the data in Ref. \cite{Kawaguchi2014} as shown in the Supplementary material Fig.\ S1.  

\subsection*{Resistivity measurements}
The temperature dependence of the electrical resistivity was measured by a  
four-contact method in a Quantum Design physical property measurement system 
(PPMS) in magnetic fields up to 14 T. Examples of the temperature dependence of 
the resistivity in zero and applied magnetic fields are shown in Supplementary material (Figs.\ S2-S7). The high-field measurements were performed in DC magnetic fields up to 35 T at the National High Magnetic Field Laboratory, Tallahassee, FL, USA. The 
high-field transport measurements in pulsed magnetic fields up to 67 T were 
performed at the Dresden High Magnetic Field Laboratory at HZDR and at the 
National High Magnetic Field Laboratory, Los Alamos, NM, USA. The superconducting transition temperature $T_{\rm c}$, as given in the 
paper, was determined using $T_{{\rm c},{90}}$ as shown in the Supplementary material (Figs.\ S6 and S7). Other criteria, such as 50\% of the normal state resistance, yield qualitatively the same temperature dependence of $H_{\rm c2}$. The SDW transition temperature $T_{\rm N}$ was defined as the peak position of the temperature derivative of the resistivity curves in analogy to the procedure applied for bulk single crystals \cite{Pratt2008}, see Supplementary material Fig.\ S2.

The measurements were performed in magnetic fields applied along the crystallographic $c$-axis of the films, which coincides with the normal direction of the films surface. Therefore, the $H_{\rm c2}$ data presented in the paper depend on the in-plane coherence
length $\xi_{\rm ab}$ only, which is unaffected by the film thickness $D_{\rm film} \sim$ 100 nm. Additionally, $\xi_c > d/2$ is satisfied for all doping levels, where $d$ is the spacing between the neighboring FeAs layers. The estimates given in the Supplementary material indicate that the fluctuation effects close to $T_{\rm c}$ can be neglected in our case. We assume that the transition width is related to small inhomogeneities in the P distribution and to a difference between $H_{\rm c2}(T)$ and $H_{\rm irr}(T)$, where $H_{\rm irr}$ is the irreversibility field. In particular, $H_{\rm irr}(T)$ is noticeably affected by flux pinning at low temperatures and high magnetic fields. Thus, our consideration of BaFe$_2$(As$_{1-x}$P$_x$)$_2$ thin films as 3D superconductors and the neglect of 2D corrections and fluctuation effects are indeed justified.

\section*{Acknowledgements}

This work was supported by DFG (GR 4667/1-1). S.-L.D, D.E., I.C. and I.M. thank the VW-foundation for financial support. D.E. also thanks RSCF-DFG Grant. The work at NHMFL was supported by the National Science Foundation Cooperative Agreement No. DMR-1157490 and the State of Florida. K.I. acknowledges the Open Partnership Joint Projects of JSPS Bilateral Joint Research Projects. We acknowledge the support of the HLD at HZDR, member of the European Magnetic Field Laboratory (EMFL). I.C. and I.M. thank the support RSF, grant No. 16-42-01100 and RFBR grant No. 15-03-99628a. We ,also, acknowledge fruitful discussions with D.\ Daghero, T.\ Terashima and J.\ Wosnitza. The publication of this article was funded by the Open Access Fund of the Leibniz Association.

\section*{Author contributions statement}

V.G., K.I. and F.K. designed the study. V.G. 
analyzed $H_{\rm c2}$ data, and wrote the manuscript. D.V.E. and S.-L.D. 
provided theoretical support in data analysis. I.C., I.M. and A.Y. prepared the 
PLD targets. Thin PLD films were prepared by V.G., K.I. and F.K. High-field 
measurements were performed by J.H., T.F., C.T., J.J., B.M., M.J., F.K., K.I. 
and V.G. V.G. performed transport measurements at magnetic fields up to 14 T. 
I.N., R.F., T.H. and H.I. prepared and characterized MBE thin films. K.I., J.H., 
H.I. and R.H. supervised the project. All authors discussed the results and 
implications and commented on the manuscript. 

\section*{Additional information}

The authors declare no competing financial interests. Correspondence should be addressed to V.G. (v.grinenko@ifw-dresden.de)

\thispagestyle{empty}

\maketitle 
\section*{Supplementary materials}
\appendix
\chapter{}

\setcounter{figure}{0} 
\renewcommand\thefigure{\thesection S\arabic{figure}}    

\maketitle

In the present Supplementary material we show the doping dependence of the $c$-axis lattice parameter for BaFe$_2$(As$_{1-x}$P$_x$)$_2$ films grown on MgO and LaAlO$_3$ (LAO) substrates, the temperature dependences of the electrical resistance in magnetic fields, the criteria used for the determination of the SDW transition temperature $T_{\rm N}$ and the superconducting critical temperature $T_{\rm c}$ in static magnetic fields as well as the upper critical field $H_{\rm c2}$ in pulsed magnetic fields. We estimated the  fluctuation effect on the superconducting transition width and that on the evaluated values of the slope of the upper critical field. Finally, we  provide tables with a list of the fitting parameters described in the main text.

\subsection*{Composition of the films}

\begin{figure}
\centering
	\includegraphics[width=9.1cm]{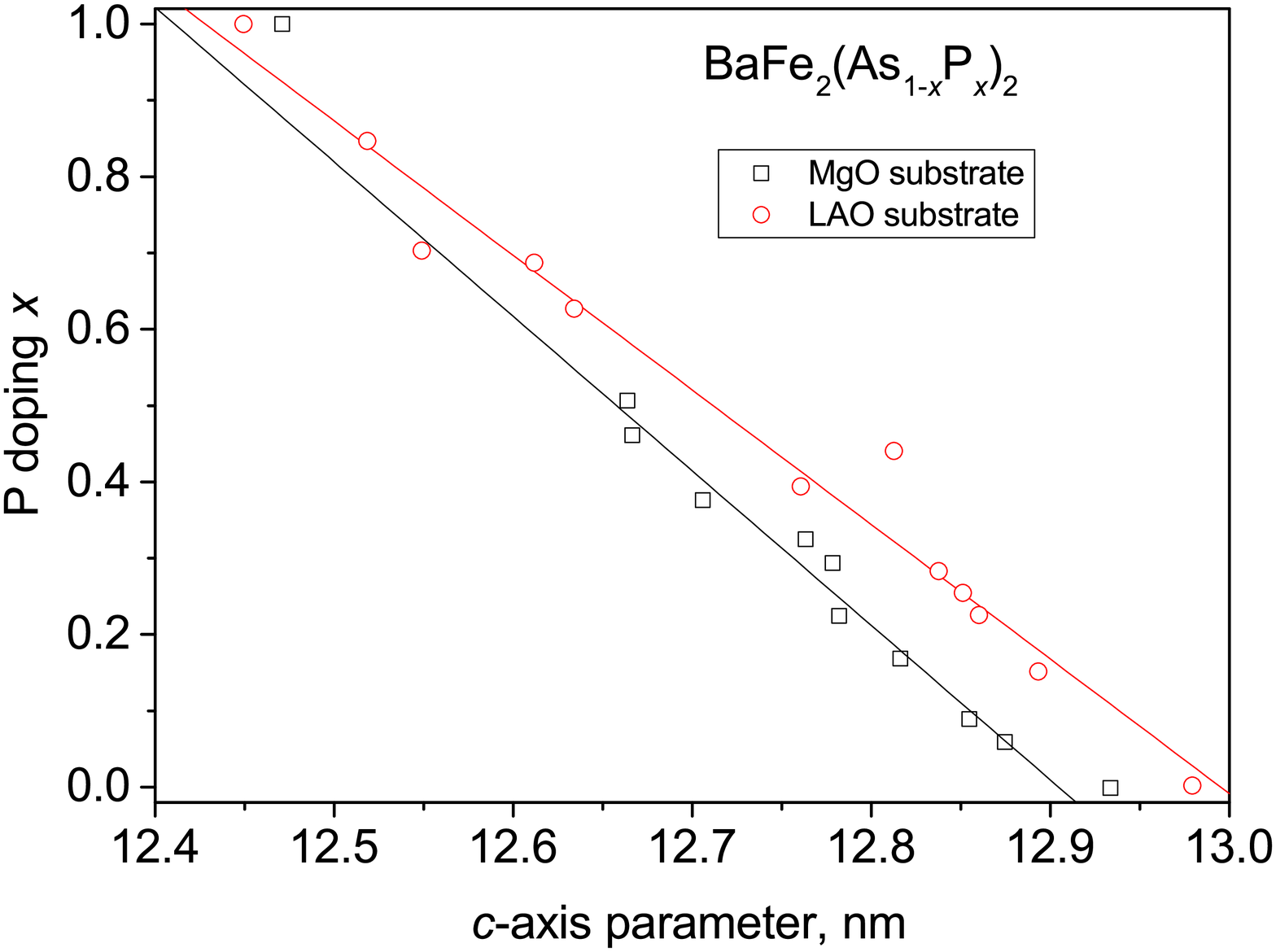}
\caption{(a) Relation between the $c$-axis lattice parameter and the P doping level $x$ of BaFe$_2$(As$_{1-x}$P$_x$)$_2$ films grown on MgO and LaAlO$_3$ (LAO) substrates. The data are taken from previous studies.\cite{Kawaguchi2014s}}
\label{cvsK}
\end{figure}

The P doping level of the thin films given in the main text were calculated using the relation between the $c$-axis lattice parameter and the P doping obtained in previous studies (Fig.\ \ref{cvsK}).\cite{Kawaguchi2014s}

\subsection*{Evaluation of $T_{\rm N}$ and $T_{\rm c}$}

\begin{figure}
\centering
	\includegraphics[width=9.1cm]{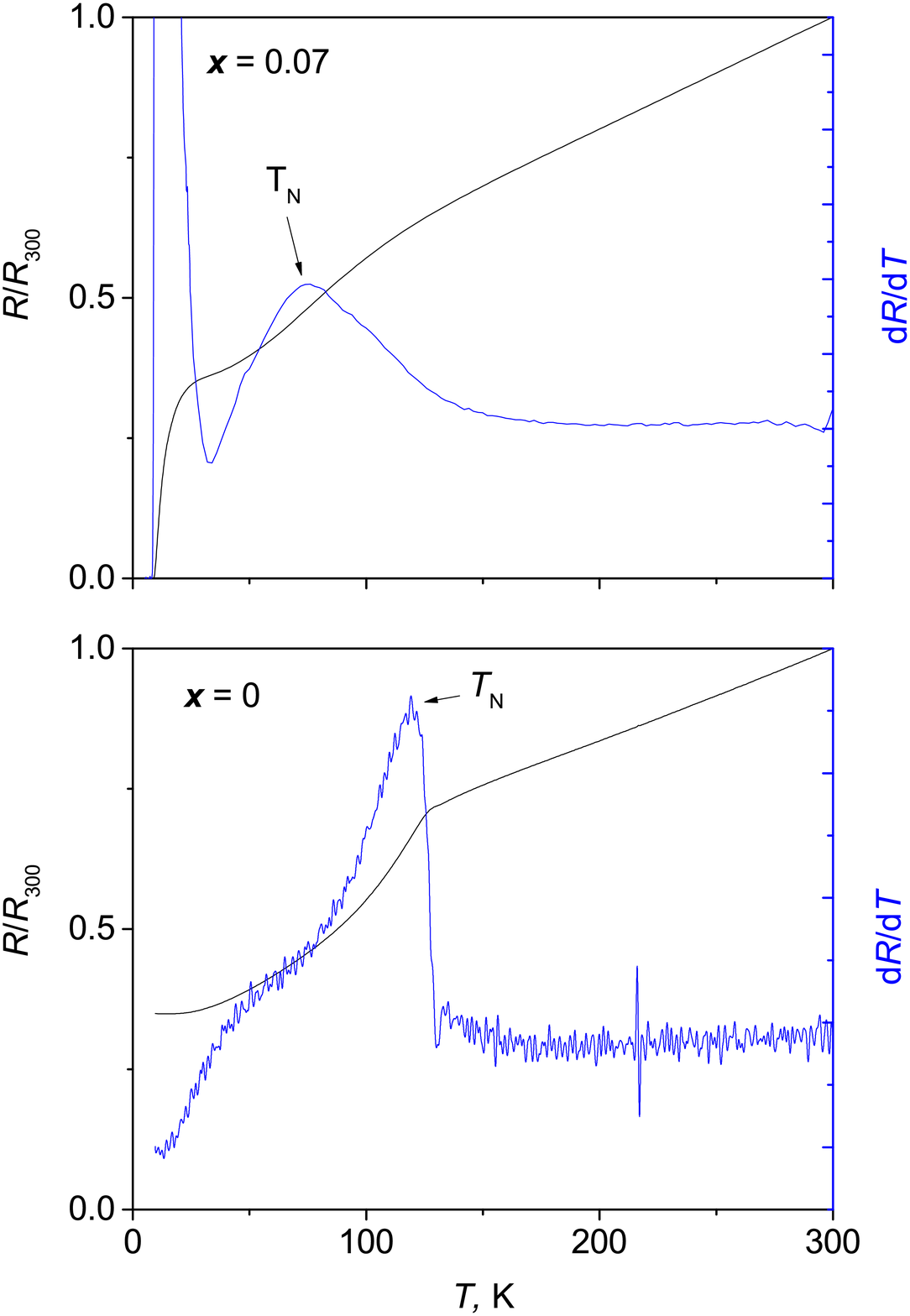}
\caption{An example of the temperature dependence of the resistance of BaFe$_2$(As$_{1-x}$P$_x$)$_2$ films with SDW transition at low temperatures (left) and its derivative (right). The peak position of the derivative is assigned as $T_{\rm N}$. The criterion is based on the comparison between the neutron scattering and transport data for the BaFe$_2$As$_2$ system.\cite{Pratt2008}} 
\label{TN}
\end{figure}

The spin density wave transition temperatures $T_{\rm N}$ of the BaFe$_2$(As$_{1-x}$P$_x$)$_2$ films were defined by a standard procedure developed for single crystals (Fig. \ref{TN}).\cite{Pratt2008s}

\begin{figure}
\centering
\includegraphics[width=26pc]{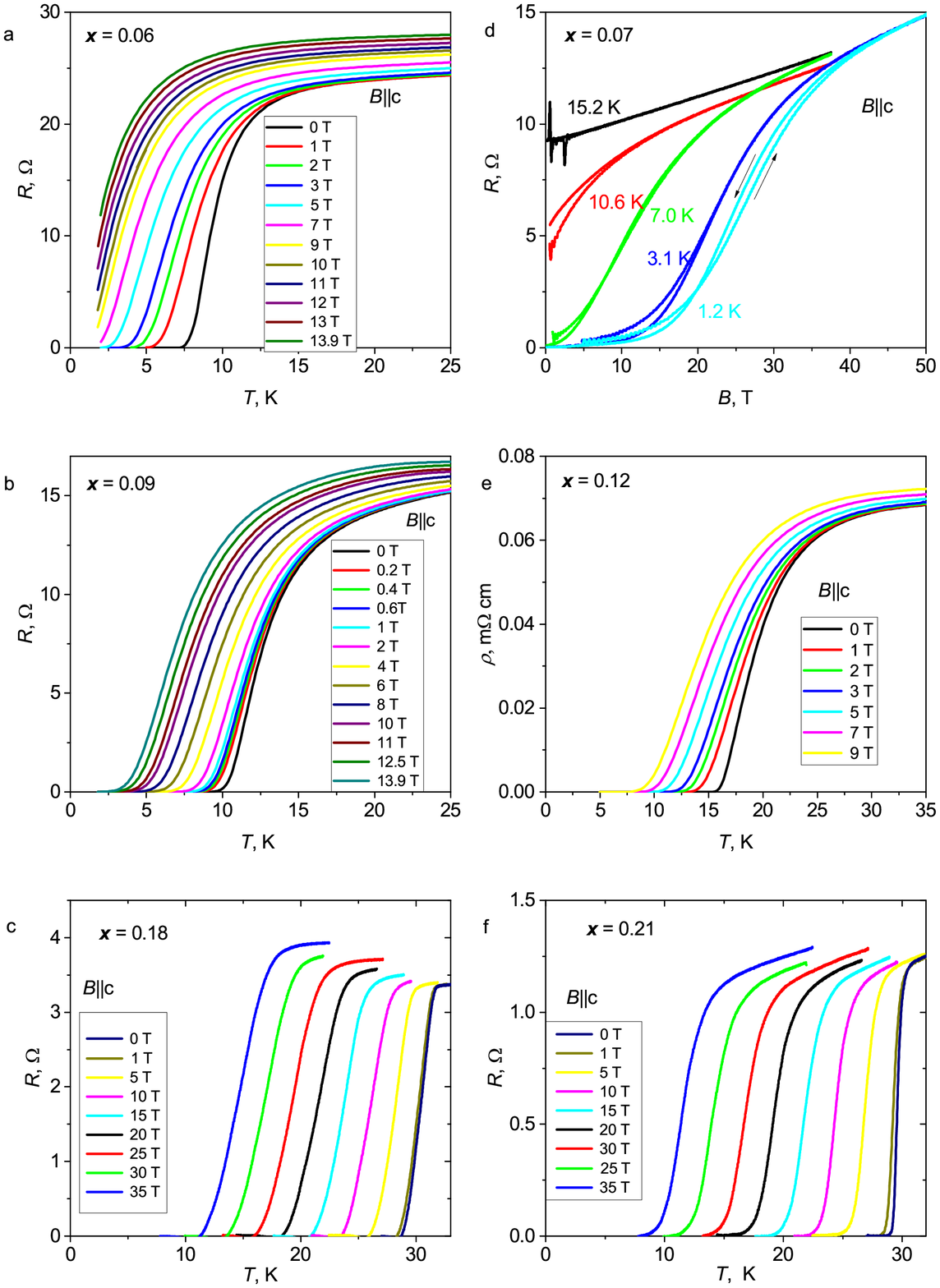}
\caption{Temperature dependences of the resistance (resistivity if available) of BaFe$_2$(As$_{1-x}$P$_x$)$_2$ films prepared by MBE in static magnetic fields and field dependences of the resistivity in pulsed magnetic fields. The doping levels, fields strength and temperatures are given in the figures too.} 
\label{RT1}
\end{figure}

\begin{figure}
\centering
\includegraphics[width=26pc]{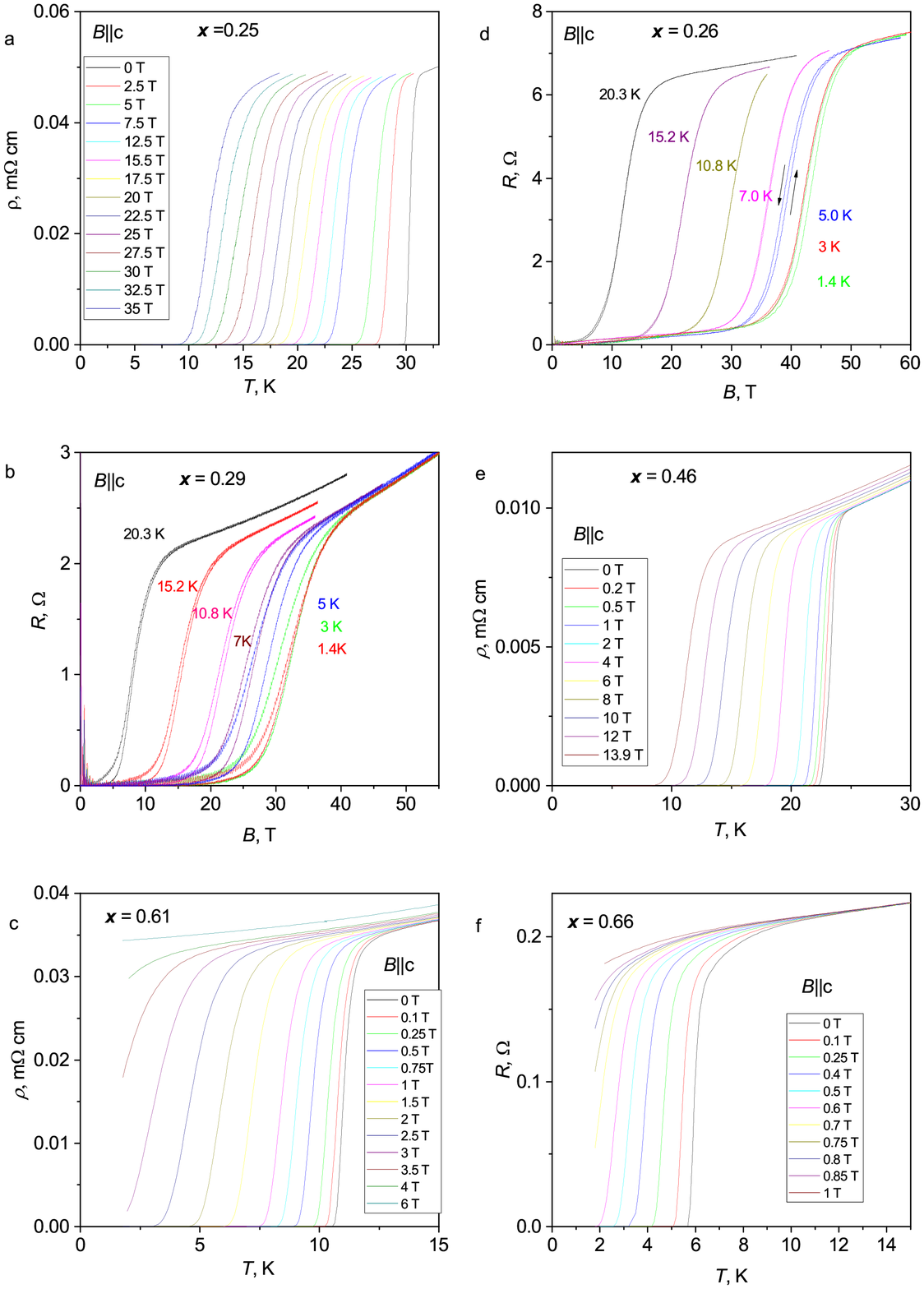}
\caption{Temperature dependences of the resistance (resistivity if available) of BaFe$_2$(As$_{1-x}$P$_x$)$_2$ films prepared by MBE in static magnetic fields and field dependences of the resistivity in pulsed magnetic fields. The doping levels, fields strength and temperatures are given in the figures too.} 
\label{RT2}
\end{figure}

\begin{figure}
\centering
\includegraphics[width=26pc]{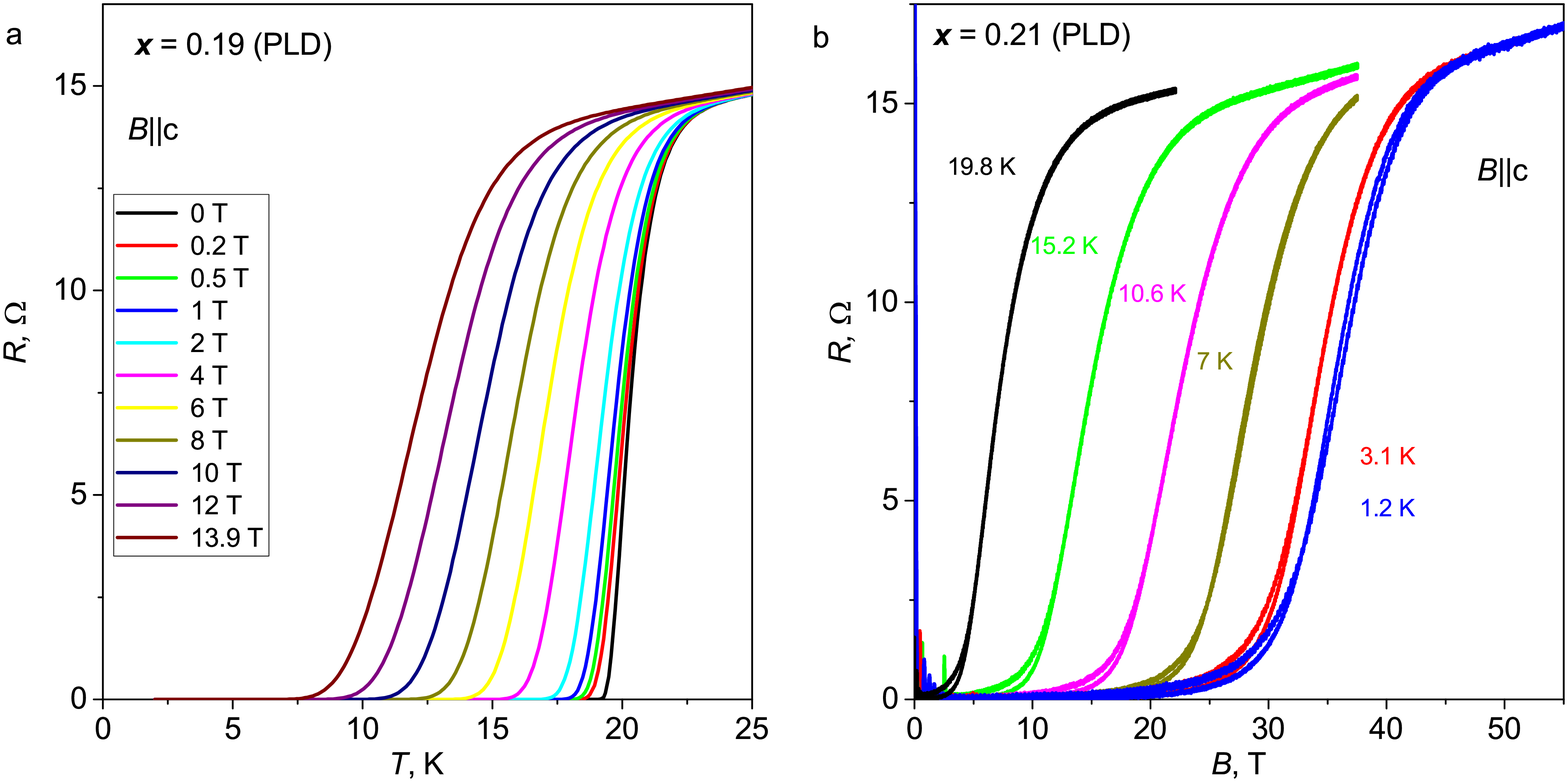}
\caption{Temperature dependences of the resistance of BaFe$_2$(As$_{1-x}$P$_x$)$_2$ films prepared by PLD in static magnetic fields and field dependences of the resistivity in pulsed magnetic fields. The doping levels, fields strength and temperatures are given in the figures too.} 
\label{RT3}
\end{figure}

The temperature dependencies of the upper critical fields $H_{\rm c2}$ given in the main text were obtained from resistivity measurements in static and pulsed magnetic fields  (Figs.\ \ref{RT1}, \ \ref{RT2}, and \ \ref{RT3}).

\begin{figure}
\centering
\includegraphics[width=28pc]{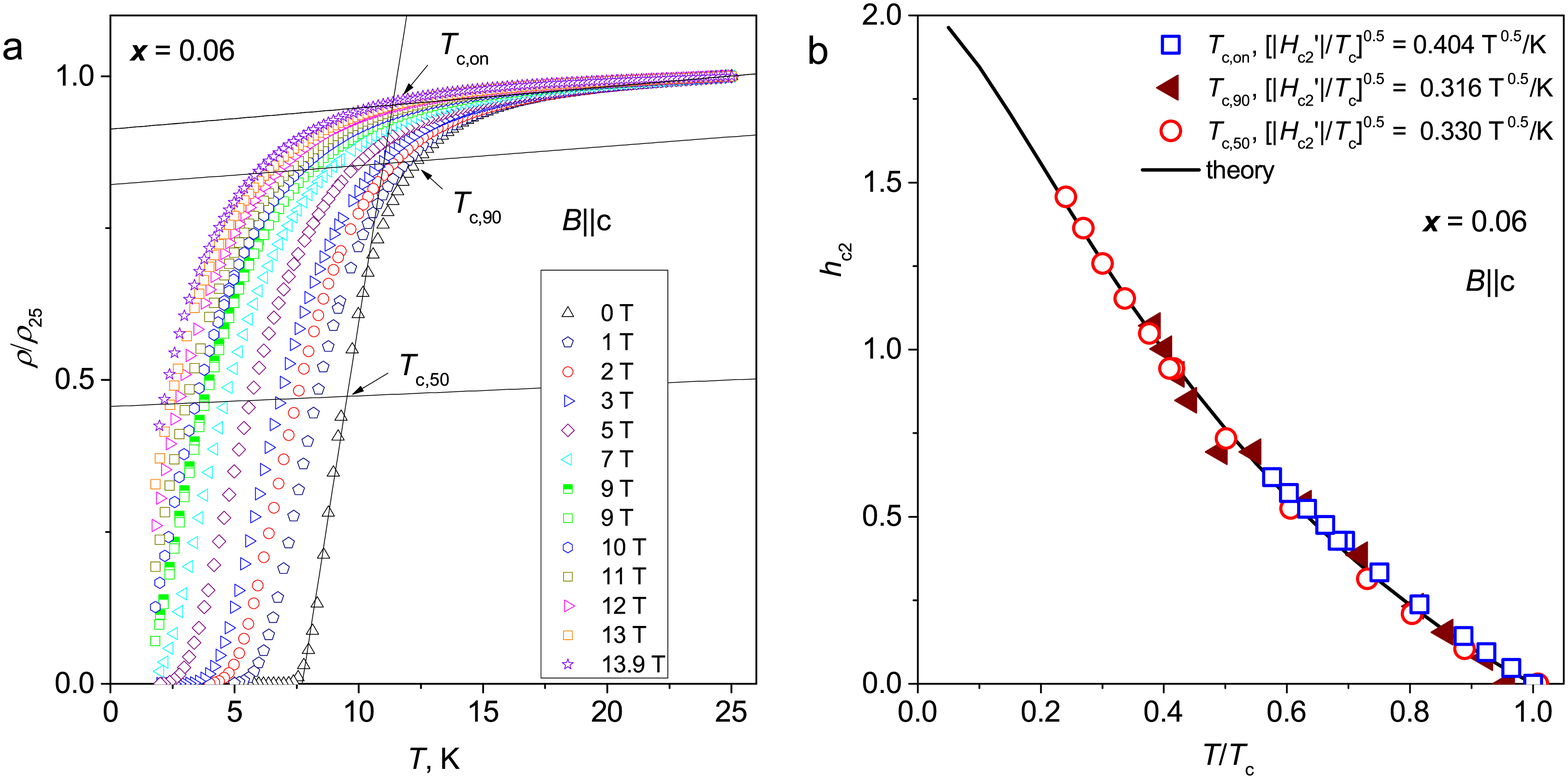}
\caption{(a) Temperature dependences of the resistance of a BaFe$_2$(As$_{0.94}$P$_{0.06}$)$_2$ film prepared by MBE in static magnetic fields. $T_{{\rm c},{\rm on}}$, $T_{{\rm c},{90}}$, and $T_{{\rm c},{50}}$ denote different criteria used to plot the $h_{\rm c2} = \frac{H_{\rm c2}}{-H_{\rm c2}'T_{\rm c}}$ values versus reduced temperature $T/T_{\rm c}$ as shown in panel (b).} 
\label{hc21}
\end{figure}

\begin{figure}
\centering
\includegraphics[width=28pc]{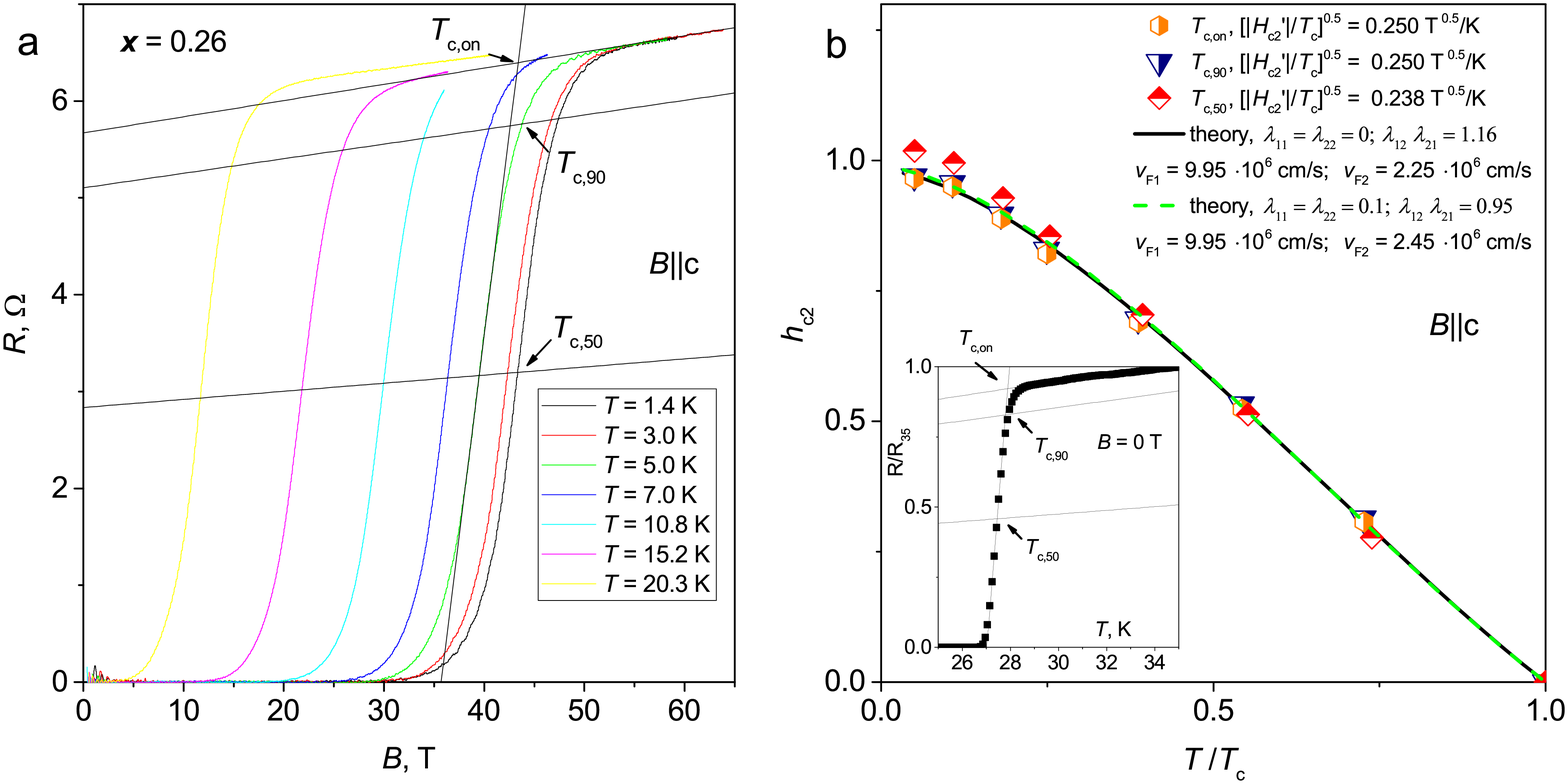}
\caption{(a) Field dependences of the resistance of a BaFe$_2$(As$_{0.74}$P$_{0.26}$)$_2$ film prepared by MBE (pulsed field measurements). $T_{{\rm c},{\rm on}}$, $T_{{\rm c},{90}}$, and $T_{{\rm c},{50}}$ denote different criteria used to plot the $h_{\rm c2} = \frac{H_{\rm c2}}{-H_{\rm c2}'T_{\rm c}}$ values versus reduced temperature $T/T_{\rm c}$ as shown in panel (b). The temperature dependence of the normalized resistance $R/R_{35}$ is shown in the inset of panel (b). The fitting curves correspond to two different cases with zero intraband coupling constants - solid line and non-zero intraband coupling constants - dased line, where $\Omega_{\rm sf} = 62$ K.} 
\label{hc22}
\end{figure}

We present two examples of the $H_{\rm c2}$ temperature dependences defined using different criteria $T_{{\rm c},{\rm on}}$, $T_{{\rm c},{90}}$, and $T_{{\rm c},{50}}$ as shown in Figs. \ref{hc21}a and \ref{hc22}. In all cases, the different criteria result in a quantitative but not a qualitative change of the $H_{\rm c2}$ dependences (Figs. \ref{hc21}b, and \ref{hc22}b). To exclude possible effects of the irreversibility field $H_{\rm irr}$ on the transition width, in the main text we used $T_{{\rm c},{90}}$ to plot $H_{\rm c2}$ (for discussion see below).   

\subsection*{Two-band model for $H_{\rm c2}$}

The doping evolution of the temperature dependencies of $H_{\rm c2}$ was described by the two-band model for a clean superconductor as proposed by Gurevich \cite{Gurevich2010s,Gurevich2011s}. Its expression for $B\parallel c$ is given by

\begin{equation}
\label{Eq1}
a_1G_1 + a_2G_2 + G_1G_2 = 0,
\end{equation}
where
 \begin{equation*}
  \begin{aligned}
G_1 = {\rm ln}t + 2e^{q^2}{\rm Re} \sum_{n=0}^\infty \int_{0}^\infty e^{-u^2}[\frac{u}{n+1/2} - \frac{t}{\sqrt{b}}{\rm tan}^{-1}(\frac{u\sqrt{b}}{t(n+1/2)+{\rm i}\alpha b})]du,
 \end{aligned}
\end{equation*}

$G_2$ is obtained by substituting $\sqrt{b} \rightarrow \sqrt{\eta b}$ 
and $q \rightarrow q\sqrt{s}$ in $G_1$, 

$a_1 = (\lambda_0 + \lambda_-)/2\omega, a_2 = (\lambda_0 - \lambda_-)/2\omega$,

$\lambda_- =\lambda_{11} - \lambda_{22},\lambda_0 = (\lambda^2_- + 4\lambda_{12}\lambda_{21})^{1/2},\omega = \lambda_{11}\lambda_{22} -\lambda_{12}\lambda_{21}$,

and $b = \frac{\hslash^2v_{{\rm F}1}^2 H}{8\pi\phi_0k_{\rm B}^2T_{\rm c}^2}$, $q^2 = \frac{Q_z^2\varepsilon_1\phi_0}{2\pi H}$, $\alpha = \frac{4\mu\phi_0T_{\rm c}}{\hslash^2v_{{\rm F}1}^2}$,

where $t = T /T_{\rm c}$, $\eta = (v_{{\rm F}2}/v_{{\rm F}1})^2$, $s = \varepsilon_2/\varepsilon_1$, 
$v_{{\rm F}i}$ is the in-plane Fermi velocity in band $i = 1, 2$, and $\varepsilon_{i} = m_{i}^{ab}/m_{i}^c$ is the mass anisotropy ratio, $\phi_0$ is the flux quantum, $\mu$ is the magnetic moment of a quasiparticle, $\lambda_{ii}$ intraband and $\lambda_{ij}$, ($j = 1, 2$, and $i\neq j$) interband pairing constants, and $\alpha \approx \alpha_{\rm M}/1.8$, where the Maki parameter $\alpha_{\rm M} = 2^{1/2} H_{\rm c2}^{\rm orb}/H_{\rm p}$ quantifies the strength of the paramagnetic pair breaking. The Fulde-Ferrel-Larkin-Ovchinnikov (FFLO) wave vector ${\bf Q}$ is determined by the condition that $H_{\rm c2}$ is at maximum, where $Q_Z$ is its projection onto $c$-axis.

\subsection*{Effect of inhomogeneities and fluctuations on the superconducting transition width} 

The underdoped films have a relatively broad transition to the superconducting state $\Delta T_{\rm c}$ (see Figs. \ref{RT1} and \ref{hc21}a). This broadening of the transition can be explained by a small inhomogeneity of the P doping within $\sim 1\%$ together with a strong doping dependency of $T_{\rm c}$ in the coexistence state between SDW and superconductivity and possible additional broadening in an applied magnetic field due to the irreversibility field $H_{\rm irr}$. For type-II superconductors in applied magnetic fields the temperature where $R = 0$ depends on the strength of vortex pinning, i.e. the condition where the critical current $I_{\rm c} = 0$ (for sufficiently small measurement currents) and not by $H_{\rm c2}$ which is related to the condition when the normal vortex cores overlap. Therefore, $R = 0$ corresponds to the irreversibility line $H_{\rm irr}$ rather than to $H_{\rm c2}$. In general, $H_{\rm irr}$ can differ considerably from $H_{\rm c2}$.\cite{Blatter1994s, Prozorov2008s} Therefore, we avoided to employ this $R = 0$ criterion. We note that the high-field data presented in Ref. \cite{Putzke2014s} reflect the measurements of $H_{\rm irr}$, which actually differs from $H_{\rm c2}$.

\begin{table}[ht]
\centering
\captionof{table}{The parameters obtained from the fit of $H_{\rm c2}$ temperature dependences shown in Fig.\ 2 main text. $\Omega_{\rm sf} = 100$ K. The crystallographic $c$-axis length is given in nm, the Fermi velocities are given in $10^6$ cm s$^{-1}$ and the transition temperature is in K.}
    \begin{tabular}{| c | c | c | c | c | c | c |}
    \hline\hline
    Substrate/technique & $c$ - axis, \AA & P - doping & $v_{\rm F1}$ &  $v_{\rm F2}$ & $[\lambda_{12}\lambda_{21}]^{0.5}$ & $T_{\rm c}$ \\ \hline
MgO/MBE & 12.877 & 0.056 & 8.4 & 0.65 & 0.43 & 11.4 \\ \hline
MgO/MBE & 12.870 & 0.070 & 6.4 & 1.60 & 0.48 & 14.0 \\ \hline
MgO/MBE & 12.815 & 0.182 & 8.5 & 4.07 & 0.78 & 31.7 \\ \hline
MgO/MBE & 12.800 & 0.212 & 9.3 & 3.90 & 0.75 & 30.3 \\ \hline
MgO/MBE & 12.782 & 0.248 & 10.1 & 3.05 & 0.76 & 30.7 \\ \hline
MgO/MBE & 12.774 & 0.263 & 9.8 & 3.05 & 0.73 & 27.9 \\ \hline
MgO/MBE & 12.761 & 0.291 & 10.6 & 3.50 & 0.69 & 26.9 \\ \hline
MgO/PLD & 12.811 & 0.190 & 6.0 & 5.80 & 0.70 & 21.6 \\ \hline
MgO/PLD & 12.800 & 0.212 & 9.3 & 3.08 & 0.75 & 24.9 \\ \hline
LAO/MBE & 12.736 & 0.457 & 11.1 & 4.15 & 0.64 & 23.7 \\ \hline
LAO/MBE & 12.649 & 0.610 & 11.4 & 6.46 & 0.44 & 11.4 \\ \hline
LAO/MBE & 12.619 & 0.663 & 11.3 & 7.90 & 0.34 & 6.3 \\ \hline
\end{tabular}
\label{tab1}
      
\end{table}

\begin{table}[ht]
\centering
\captionof{table}{The parameters obtained from the fit of $H_{\rm c2}$ temperature dependences shown in Fig.\ 2 main text. $\Omega_{\rm sf} = 62$ K. The crystallographic $c$-axis length is given in nm, the Fermi velocities are given in $10^6$ cm s$^{-1}$ and the transition temperature is in K.}
    \begin{tabular}{| c | c | c | c | c | c | c |}
    \hline\hline
    Substrate/technique & $c$ - axis, \AA & P - doping & $v_{\rm F1}$ &  $v_{\rm F2}$ & $[\lambda_{12}\lambda_{21}]^{0.5}$ & $T_{\rm c}$ \\ \hline
MgO/MBE & 12.877 & 0.056 & 8.20 & 0.61 & 0.55 & 11.4 \\ \hline
MgO/MBE & 12.870 & 0.070 & 6.50 & 1.37 & 0.62 & 14.0 \\ \hline
MgO/MBE & 12.815 & 0.182 & 8.50 & 3.80 & 1.25 & 31.7 \\ \hline
MgO/MBE & 12.800 & 0.212 & 9.70 & 3.00 & 1.18 & 30.3 \\ \hline
MgO/MBE & 12.782 & 0.248 & 10.20 & 2.20 & 1.20 & 30.7 \\ \hline
MgO/MBE & 12.774 & 0.263 & 9.95 & 2.25 & 1.08 & 27.9 \\ \hline
MgO/MBE & 12.761 & 0.291 & 11.05 & 2.45 & 1.03 & 26.9 \\ \hline
MgO/PLD & 12.811 & 0.190 & 6.00 & 5.80 & 1.20 & 21.6 \\ \hline
MgO/PLD & 12.800 & 0.212 & 9.60 & 2.60 & 0.96 & 24.9 \\ \hline
LAO/MBE & 12.736 & 0.457 & 11.30 & 3.55 & 0.91 & 23.7 \\ \hline
LAO/MBE & 12.649 & 0.610 & 11.60 & 6.23 & 0.55 & 11.4 \\ \hline
LAO/MBE & 12.619 & 0.663 & 11.40 & 7.80 & 0.41 & 6.3 \\ \hline
\end{tabular}
\label{tab2}
      
\end{table}
\vspace{0.2cm}

The effect of superconducting fluctuations plays a secondary role for the $\Delta T_{\rm c}$ values of our thin films. For example, the films with the highest $T_{\rm c} \sim 30$ K have a transition width of about 1 K, only. Note that, even these transitions width are not yet dominated by fluctuations. Quantitatively, one can estimate the temperature range where fluctuations play a role using the Ginzburg parameter defined in 3D (see also Methods section in the main text) as $\Delta T_{\rm c}/T_{\rm c} = Gi \approx 80 (T_{\rm c}/E_{\rm F})^4 \sim 10^{-4} - 10^{-5}$.\cite{Larkin} Taking $T_{\rm c} = 30$ K and the largest $E_{\rm F} \sim 100$ meV in our system according to ARPES data.\cite{Yoshida2011s} Alternatively, the $Gi$ parameter can be estimated directly from the superconducting parameters: $\Delta T_{\rm c}/T_{\rm c} = Gi = (\Gamma k_{\rm B} T_{\rm c}/H_{\rm cm}(0)^2\xi_0^3)^2/2$.\cite{Blatter1994s, Chaparro2012s} Taking the experimental values for  $T_{\rm c} = 30$ K, $H_{\rm c2}$ = 470 kG, $H_{\rm c1}$ = 600 G (field along crystallographic $c$-axis),\cite{Putzke2014s} $\lambda \approx 3 \cdotp 10^{-5}$ cm, $\xi_{\rm c} = [\Phi_0/2\pi H_{\rm c2}]^{0.5} \approx 3 \cdotp 10^{-7}$ cm and the anisotropy $\Gamma \sim$ 2 - 4, one arrives at the same estimates. However, close to $T_{\rm c}$ the $c$-axis coherence length $\xi_{\rm c}(T)$ diverges as $(1-T/T_{\rm c})^{-0.5}$. Therefore, in close vicinity to $T_{\rm c}$, where $\xi_{\rm c}(T) > D_{\rm film}$, one can consider the films as 2D superconductors, where $D_{\rm film} \approx 100$ nm is the films thickness. For optimally doped films with high $H_{\rm c2}$ values one can estimate that $\xi_{\rm c}(T) \sim D_{\rm film}$ holds only for a very narrow temperature range of $\Delta T/T_{\rm c} \approx 0.001$. On the other hand, this range is about 0.5 K for overdoped films with low $T_{\rm c} \sim 10$ K and small $H_{\rm c2} \sim 10$ kG. However, due to the low $T_{\rm c}$, the fluctuation effect is rather weak $\Delta T_{\rm c}/T_{\rm c} = Gi \approx (T_{\rm c}/E_{\rm F}) \sim 10^{-3}$ even in 2D case.

Finally we list parameters obtained from the fits of $H_{\rm c2}$ temperature dependences shown in the main text (Tabs. \ref{tab1}, and \ref{tab2}), for the case of a zero intraband coupling $\lambda_{11} = \lambda_{22} = 0$.

\end{document}